\documentclass[nolinenumbers,trackchanges,twocolumn]{aastex7}

\newcommand\aastex{AAS\TeX}

 \usepackage{amsmath}
     \usepackage{xcolor}
    \usepackage{soul}
\begin{document}

\title{Positive superhumps in high mass ratio cataclysmic variables driven by apsidal disk precession}

\author[0000-0003-2401-7168]{Rebecca G. Martin}
\affiliation{Department of Physics and Astronomy, University of Nevada, Las Vegas, 4505 South Maryland Parkway, Las Vegas, NV 89154, USA}
\affiliation{Nevada Center for Astrophysics, University of Nevada, Las Vegas, 4505 South Maryland Parkway, Las Vegas, NV 89154, USA}
\email{rebecca.martin@unlv.edu}

\author[0000-0002-4636-7348]{Stephen H. Lubow}
\affiliation{Space Telescope Science Institute, 3700 San Martin Drive, Baltimore, MD 21218, USA}
\email{}

\author[0000-0003-4497-2680]{David Vallet}
\affiliation{Department of Physics and Astronomy, University of Nevada, Las Vegas, 4505 South Maryland Parkway, Las Vegas, NV 89154, USA}
\affiliation{Nevada Center for Astrophysics, University of Nevada, Las Vegas, 4505 South Maryland Parkway, Las Vegas, NV 89154, USA}
\email[]{}

\author[0000-0003-2270-1310]{Stephen Lepp}
\affiliation{Department of Physics and Astronomy, University of Nevada, Las Vegas, 4505 South Maryland Parkway, Las Vegas, NV 89154, USA}
\affiliation{Nevada Center for Astrophysics, University of Nevada, Las Vegas, 4505 South Maryland Parkway, Las Vegas, NV 89154, USA}
\email{}

\begin{abstract}
Previously it has been assumed that a cataclysmic variable (CV) disk can only become eccentric and display superhumps if the 3:1 resonance is located within the disk. This requires the binary mass ratio to be $q=M_2/M_1\lesssim 0.33$, where $M_1$ is the mass of the white dwarf and $M_2$ is the mass of the companion star. However, several systems with higher mass ratios have been observed to exhibit positive superhumps, posing a challenge to this picture. We present the first 3D hydrodynamic simulations to show that
eccentricity growth can occur in CV disks even when the resonance radius lies outside the disk. The finite width of the 3:1 resonance extends into the outer parts of the disk and drives eccentricity. While high mass ratio CVs more commonly show negative superhumps, our linear analysis reveals that the direction of apsidal precession is highly sensitive to the disk outer radius and surface density distribution.
Smoothed particle hydrodynamic simulations tend to suppress eccentricity gradients and favor prograde precession. These results provide a natural explanation for positive superhumps in high mass ratio CVs and show that disk structure, rather than the resonance location alone, controls the emergence of superhumps.
\end{abstract}


\section{Introduction}

Cataclysmic variables (CVs) contain a white dwarf that is accreting material from a red dwarf binary companion that fills its Roche lobe \citep[e.g.][]{Warner1995}.
Positive superhumps are periodic changes to the lightcurve of a CV on a timescale slightly longer than the orbital period. They most often occur in the low mass ratio SU Uma type systems during superoutbursts \citep{Osaki2014}. Positive superhumps are widely understood to be driven by prograde apsidal precession of an eccentric disk \citep{Whitehurst1988, Hirose1990,Lubow1991,Ichikawa1993,Murray1996,Murray1998,Simpson1998,Smith2007,Oyang2021}. The disk eccentricity grows when the disk is in contact with the 3:1 resonance through a mode coupling process \citep{Lubow1991b}. 

The radius of the 3:1 resonance is outside of the tidal truncation radius of the disk if the binary mass ratio is $q>q_{\rm crit} \approx 0.33$ \citep{Hirose1990,Whitehurst1991,Patterson2005}. Therefore, it has long been assumed that eccentricity growth only occurs in CVs with $q\lesssim 0.33$, although the critical value has been widely debated and even suggested to be as low as 0.22 for a cold disk \citep{Smak2020}.
There have been several observations of positive superhumps in systems with large mass ratios, see Table~\ref{table}. While mass ratios are observationally difficult to constrain, here we list CVs that have  mass ratios that may be larger than the critical and have shown positive superhumps. This appears to be a contradiction to the theory \citep{Bruch2023, Bruch2023b}.

\begin{table*}
\centering
\begin{tabular}{llllllc}
\hline
System & Type & $P_{\rm b}$ (d) & $P_{\rm psh}$ (d) & $\epsilon$ & $q$  & References \\
\hline
MV Lyrae & NL (VY Scl) &  0.133 & 0.138 & 0.036 & $0.43^{+0.19}_{-0.13}$ & (1) \\
U Gem & DN (U Gem)  & 0.177 & 0.2 & 0.13 & $0.364 \pm 0.017$ & (2),(3)\\
AT Cnc & DN (Z-Cam) & 0.202 & 0.227 & 0.125 & 0.511 & (4),(5),(6)\\
CN\,Vel & NL (Old nova) & 0.22338 & 0.24513 & 0.097 & $>0.39$ & (7) \\
TV\,Col & DN (IP) & 0.2286 & 0.264 & 0.155 & 0.62-0.93  & (8),(9) \\
KIC\,9406652 & NL (IW~And) & 0.254509 & 0.29071 & 0.1422 & $0.83\pm0.08$ & (10),(11) \\
EI\,UMa & NL (IP candidate) & 0.2684 & 0.317-0.354 & 0.181-0.319 & $>0.48$  & (7) \\
RZ\,Gru & NL (UX\,UMa) & 0.41750 & 0.520 & 0.2455 & $>0.48$ & (10) \\
\hline
\end{tabular}
\caption{CVs with reported  positive superhumps and high mass ratios. The orbital period of the binary is $P_{\rm b}$. The period of the positive superhump is $P_{\rm psh}$. The period excess is $\epsilon = (P_{\rm psh}-P_{\rm b})/P_{\rm b}$ and the binary mass ratio is $q=M_2/M_1$. Mass ratio lower limits have been estimated with the secondary mass-period relation of \cite{Knigge2011} \citep[see][]{Bruch2023b}. References: (1) \cite{Skillman1995} (2) \cite{Smak2001} (3) \cite{Smak2004} (4) \cite{Sun2025} (5) \cite{Nogami1999} (6) \cite{Bruch2019} (7) \cite{Bruch2023b}  (8) \cite{Retter2003} (9) \cite{Hellier1993} (10) \cite{Bruch2023} (11) \cite{Gies2013}. NL = nova like. DN = dwarf nova. IP = intermediate polar. }
\label{table}
\end{table*}

The 3:1 resonance has a nonzero width that allows the effects of the resonance to extend into the outer parts of the disk even when the resonance itself lies outside of the tidal truncation radius of the disk \citep[e.g.][]{Lubow1991,Ogilvie07,Padgett2026}.
Therefore, the effects of the 3:1 resonance may remain significant even in high mass ratio systems.
Linear analyses \citep[e.g.][]{GO06, Ogilvie2008} provide a means of exploring the effects
of the 3:1 resonance width \citep{Lubow2010, Vallet2026, Lubow2026}.  
For a steady state disk profile, this sometimes leads to retrograde apsidal precession. This retrograde precession can explain the negative superhumps that are most often observed for high mass ratio binaries.  \cite{Kley2008} reported eccentricity growth for all binary mass ratios with the fastest growth rate with $q \approx 0.3$ in a grid based 2D simulation.  They did not pursue an explanation for this effect but speculated that it might involve higher order non-linear mode couplings. We suggest that the effects of the resonance width play an important role.

In this Letter, we demonstrate that eccentricity growth can occur even when the 3:1 resonance lies outside the disk, and show that the direction of apsidal precession is controlled by the disk structure.  This may explain the positive superhumps observed in high mass ratio binaries that has been a puzzle for some time. 
In Section~\ref{sec:hydro} we present the first 3D hydrodynamical simulations that show eccentricity growth for mass ratios above the critical value.  In Section~\ref{analytic} we use linear analysis to explore how the truncation radius of the disk and the density distribution can affect the direction of the apsidal precession and allow for positive superhumps even in systems with a high mass ratio. Furthermore, we explain why SPH simulations generally show only prograde apsidal precession.   We draw our conclusions in Section~\ref{concs}.

\section{Hydrodynamic simulations}
\label{sec:hydro}

In this Section we consider hydrodynamic simulations using the smoothed particle hydrodynamics (SPH) code {\sc Phantom} \citep{Price2018}. The binary consists of a white dwarf with mass $M_1$ and a companion star of mass $M_2$ in a circular orbit  with semi-major axis $a_{\rm b}$. The orbital period of the binary is $P_{\rm b}=2\pi/\Omega_{\rm b}$, where $\Omega_{\rm b}=\sqrt{G(M_1+M_2)/a_{\rm b}^3}$. We consider binary mass ratios $q=M_2/M_1=0.1$, 0.2, 0.4 and 0.6 spanning both below and above the critical value. 
The binary components are modelled as sink particles with accretion radii of $r_1=0.05\,r_{\rm l}$ and $r_2=0.05\,a_{\rm b}$ for the white dwarf and the companion star respectively.  The Roche lobe radius is
\begin{equation}
    r_{\rm l}=\frac{0.49\,q'^{2/3}a_{\rm b}}{0.6q'^{2/3}+{\rm ln}(1+q'^{1/3})},
\end{equation}
and $q'=1/q=M_1/M_2$ \citep{Eggleton1983}. Particles that move inside of the sink radius are accreted and their mass and angular momentum are added to the sink particle \citep{Bate1995}.

The simulations begin with an accretion disk around the white dwarf that extends initially from an inner radius of $r_{\rm in}=0.05\,r_{\rm l}$ up to the outer radius of $r_{\rm out}=r_{\rm l}$ with a density distribution that scales with $\Sigma \propto r^{-1/2}$. The total disk mass is initially $M_{\rm d}=10^{-6}\,\rm M$, where $M=M_1+M_2$ is the total mass of the binary, and is spread over $500,000$ SPH particles. The disk aspect ratio is constant with radius and we take $H/r=0.03$. We choose a relatively high disk aspect ratio to maximize the resonance width and therefore increase eccentricity growth rate  for higher binary mass ratios. The \cite{Shakura1973} viscosity parameter is $\alpha=0.03$.  The disk viscosity is implemented by adapting the SPH artificial viscosity according to the procedure described in \cite{Lodato2010}, using $\alpha_{\rm AV} = 0.55$ and $\beta_{\rm AV} = 2.0$. The disk is resolved with shell-averaged smoothing length per scale height $\left<h\right>/H = 0.55$.

\begin{figure*}
    \centering
      \includegraphics[width=0.49\linewidth]{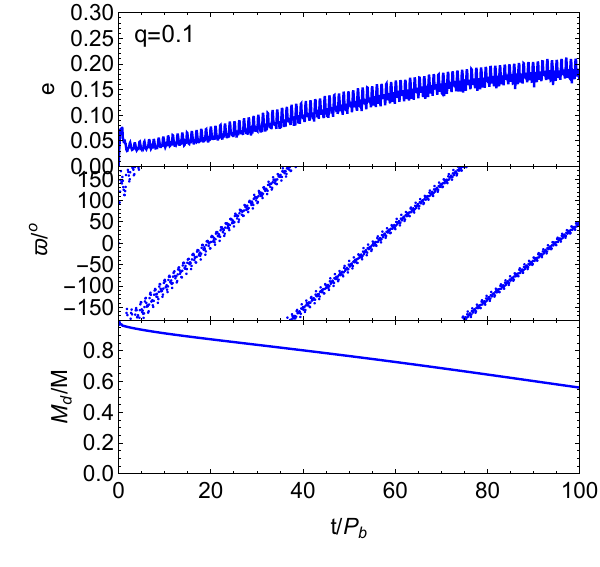}
        \includegraphics[width=0.49\linewidth]{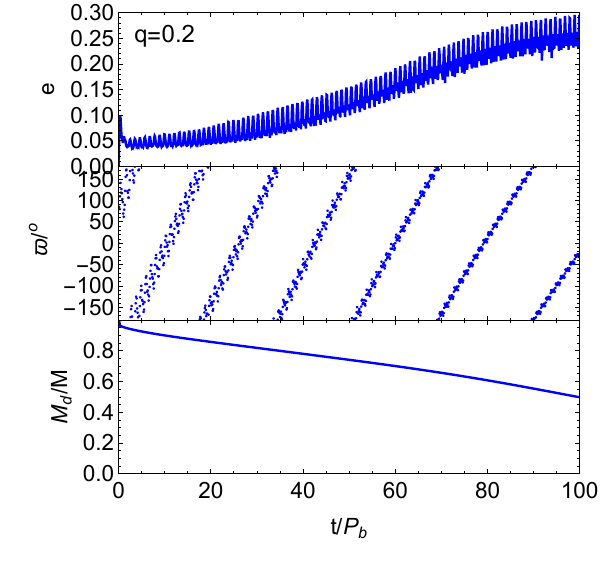}
    \includegraphics[width=0.49\linewidth]{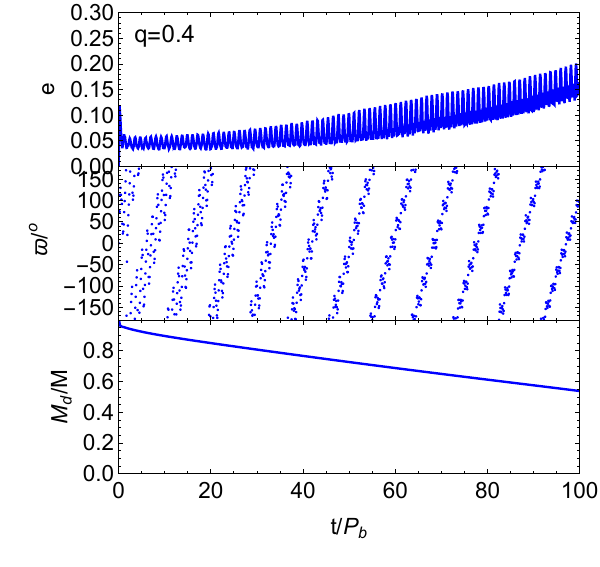}
        \includegraphics[width=0.49\linewidth]{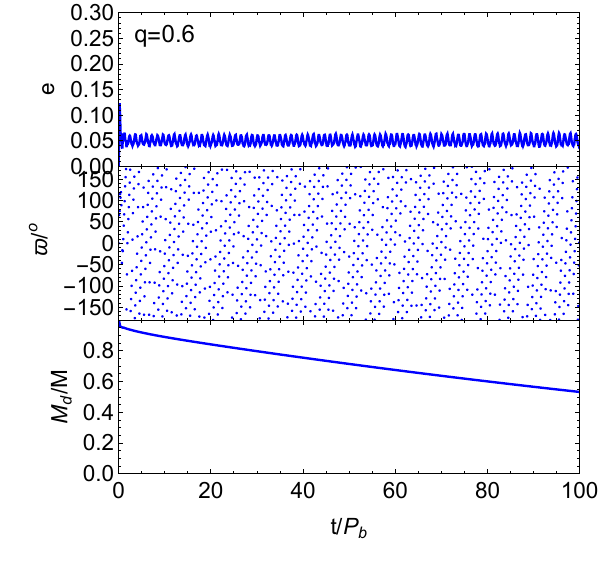}
    \caption{Results of the SPH simulations. The disk eccentricity, longitude of periapsis and total disk mass as a function time at a radius of $r=0.6\,r_{\rm l}$ for SPH simulations with $q=0.1$ (top left), $q=0.2$ (top right), $q=0.4$ (bottom left) and $q=0.6$ (bottom right).}
    \label{fig:sph}
\end{figure*}

Fig.~\ref{fig:sph} shows the results of the four SPH simulations with varying binary mass ratio. The upper row of each panel shows the eccentricity of the disk in time. The large eccentricity growth is unsurprising for low mass ratios ($q=0.1$ and $q=0.2$) as similar simulations have been presented previously \citep[e.g.][]{Lubow1991,Franchini2019}. However, the simulation with $q=0.4$ is the first simulation that has been presented that shows eccentricity growth in a 3D hydrodynamic simulation with a mass ratio greater than the critical, $q>q_{\rm crit}$.

The simulation with $q=0.6$ shows only very slow eccentricity growth in the outer parts of the disk. At a radius of $r=0.6\,r_{\rm l}$, the time averaged eccentricity grows from 0.050 to 0.051 between $t=50\,P_{\rm b}$ and $t=90\,P_{\rm b}$. This suggests that while eccentricity growth is possible at higher mass ratios, the growth rate may become very small and sensitive to disk properties. Since the disk mass is heavily depleted by the end of the simulation, it is not possible to show any significant growth for this case with the parameters that we have chosen.
The large viscosity in the disk leads to a fast dissipation of the disk and limits the simulation time. 
Injection of material into the disk from the Roche lobe overflow stream may allow eccentricity growth to occur even for higher mass ratios.

The middle panels of each plot in Fig.~\ref{fig:sph} show the longitude of periapsis for the disk. The apsidal precession is in a positive direction for all of the simulations. The precession timescale is shorter for a higher mass ratio binary.  These simulations suggest that prograde apsidal precession of an eccentric disk is a likely outcome, even for high mass ratio binaries. Therefore, we suggest that the positive superhumps that have been observed in high mass ratio binaries may be explained with the same mechanism that is well understood to operate for the positive superhumps observed in low mass ratio binaries.

\section{Why do the SPH simulations all show prograde precession?}
\label{analytic}

In observations of high mass ratio binaries, negative superhumps are more often seen than positive superhumps. However, the SPH simulations in the previous section imply that positive superhumps are more likely to occur for all mass ratios. In this Section we discuss why SPH simulations always show prograde precession.

\subsection{The density distribution in the SPH simulations}

\begin{figure*}
    \centering
    \includegraphics[width=0.49\linewidth]{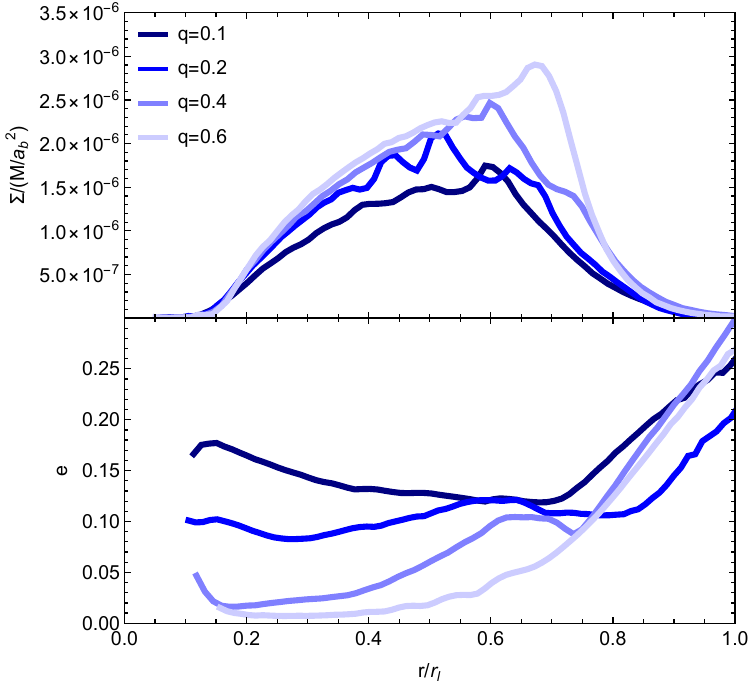}
        \includegraphics[width=0.49\linewidth]{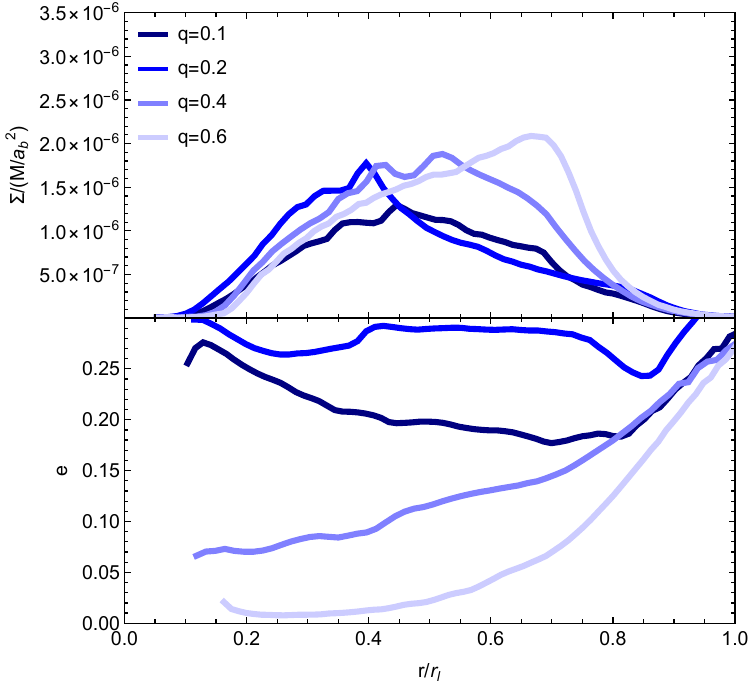}
    \caption{The disk surface density (upper) and eccentricity (lower) as a function of radius at a time of $t=50\,P_{\rm b}$ (left) and $t=100\,P_{\rm b}$ (right) for the SPH simulations. }
    \label{fig:sigma}
\end{figure*}

Fig.~\ref{fig:sigma} shows the surface density and eccentricity for the simulations at a time of $t=50\,P_{\rm b}$ and $t=100\,P_{\rm b}$. The surface density of each simulation (upper panels) shows a disk with a relatively large inner hole, as is often seen in SPH simulations \citep[e.g.][]{Lodato2004}. This leads to a radially narrow disk, partially as a result of the size of the sink particle representing the white dwarf. The communication timescale across the radial extent of the disk is relatively short, and the eccentricity profile is fairly flat with radius (lower panels). 

The direction of apsidal precession for a disk depends upon the balance between terms that involve the pressure, the radial eccentricity gradient and the dynamical gravitational driving. The pressure can drive a small prograde or retrograde component \citep{Ogilvie2008}. The radial eccentricity gradient can drive retrograde precession while the dynamical gravitational term drives prograde precession \citep[see equation 36 in][]{GO06}. The term that can dominate, and drive overall  retrograde apsidal precession (for relatively cool disks) is the term involving the radial eccentricity gradient. Since the SPH simulations are radially narrow, this unphysical surface density profile may only allow for prograde apsidal precession. This is a combination of the high dissipation which removes the eccentricity gradients and the radially narrow disk.

Some grid code simulations have shown density profiles that extend smoothly out to the Roche lobe radius at apastron \citep[see Fig. 4 in ][]{Jordan2024}. In general, the density profiles can be closer to a power law with radius \citep[e.g. see Fig. 13 in][]{Kley2008}. In these grid code simulations, retrograde precession is often observed, in contrast to the SPH simulations. We suggest that the density profile in the SPH simulations is a primary factor why they all show positive superhumps.
The temperature structure of the disk can also play a role and we have used a fixed aspect ratio while the grid code simulations use a more sophisticated equation of state.

These SPH results have implications in particular for the possibility of superhumps in magnetic CVs. In an intermediate polar (IP), the white dwarf has a magnetic field that is strong enough to truncate the inner parts of the disk, but allow the outer disk to form.  This is somewhat similar to the SPH simulations that are presented here and therefore positive superhumps are more likely in the  narrow disk expected in intermediate polars. This could help to explain the positive superhumps observed in TV Col and EI UMa (see Table~\ref{table}).

\subsection{Effect of the outer disk radius}

\begin{figure*}
    \centering
    \includegraphics[width=0.49\linewidth]{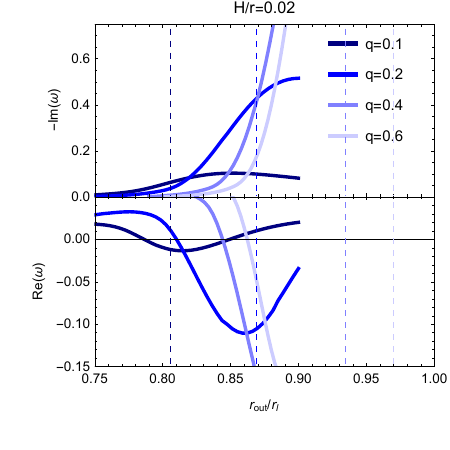}
        \includegraphics[width=0.49\linewidth]{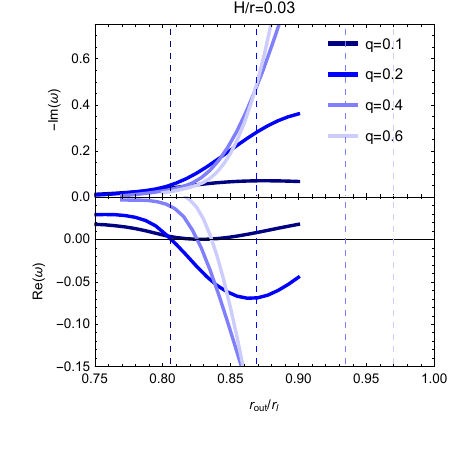}
    \caption{The eccentricity growth rate (upper) and the precession rate (lower) as a function of the disk outer radius for the linear models with surface density profile $\Sigma \propto r^{-1/2}$. The left panel shows $H/r=0.02$ and the right panels shows $H/r=0.03$. The vertical dashed lines show the 3:1 resonance location for each of the four mass ratios.  }
    \label{fig:analytic}
\end{figure*}

We use linear analysis to consider the effect of the density distribution on the superhump behavior in high mass ratio CVs. We follow the methods of \cite{Vallet2026} and \cite{Lubow2010}, originally based on \cite{GO06} but use the 3D equations \citep{Teyssandier2017} to find the complex eccentricity $E=e \exp(i \varpi)$, where $e$ is the magnitude of the eccentricity and $\varpi$ is the periapse angle. Material in the accretion disk  orbits about the white dwarf at the Keplerian frequency  given by $\Omega = \sqrt{ G M_1 /r^3}$. The disk extends from inner radius $r_{\rm in}=0.01\,a_{\rm b}$ to outer radius $r_{\rm out}$ with surface density $\Sigma\propto r^{-1/2}$, which is assumed to be fixed in time. The pressure is given by $P = c_{\rm s}^2 \Sigma$, where $H/r$ is the disk aspect ratio that is taken to be constant with radius and the sound speed  is $c_{\rm s}=(H/r) r \Omega$.

The eccentricity evolution of an adiabatic disk satisfies
\begin{equation}
J\frac{\partial E}{\partial t} = i \frac{\partial }{\partial r}\left(a \frac{\partial E}{\partial r} \right) + ibE + JsE, \label{eq:EQ1}
\end{equation}
where the angular momentum per unit radius divided by $\pi$ is given by
\begin{align}
J = 2 r^3 \Omega \Sigma.
\end{align}
We consider the 3D versions of these equations for which
\begin{align}
    a = 
    \left(2 - \frac{1}{\gamma} -i\alpha_{\rm b}
    \right)P r^3
\end{align}
and
\begin{align}
    b = \left(4 - \frac{3}{\gamma}\right) r^2 \frac{dP}{dr} + 3\left(1 + \frac{1}{\gamma}\right)Pr + J\dot{\varpi}_g
\end{align}
\citep[e.g.][]{Ogilvie2014,Teyssandier2017}, where $\gamma = 3/5$ is the gas adiabatic index and $\alpha_{\rm b}=0.01$  is an eccentricity damping parameter related to the bulk viscosity of the disk (rather than the usual shear viscosity parameter of \cite{Shakura1973}).
A free particle on an eccentric orbit undergoes gravitational precession with a rate given by
\begin{align}
    \dot{\varpi}_g = \frac{1}{4} q \left( \frac{r}{a_{\rm b}} \right)^2 \Omega \, b^{(1)}_{3/2}{ \left(\frac{r}{a_{\rm b}} \right) },
\end{align}
where 
\begin{align}
b^{(1)}_{3/2}{ \left(\frac{r}{a_{\rm b}} \right) } =  \frac{1}{\pi}\int^{2\pi}_0 \frac{ \cos{( x)} ~dx}{ \left( 1+ (r/a_{\rm b})^2 -2(r/a_{\rm b})\cos{x}   \right)^{3/2}   }
\end{align}
is the Laplace coefficient.

The radius of the 3:1 resonance is 
\begin{equation}
r_{\rm res} = 3^{-2/3}(1+q)^{-1/3} a_{\rm b}.
\label{rres}
\end{equation}
The width of the resonance due to gas pressure is 
\begin{equation}
    w_{\rm res} \simeq (H/r)^{2/3}r_{\rm res}
\end{equation}  \citep{Meyer87}. The forcing term $s$ in equation~(\ref{eq:EQ1}) acts on the resonance location and is given by
\begin{align}
    s = \chi ~\frac{ \exp{[ -(r-r_{\rm res})^2/w_{\rm res}^2  ]} }{ \sqrt{\pi}w_{\rm res} }   
\label{eq:forcing}
\end{align}
\citep{Lubow2010}\footnote{Note that there is typo in equation (9) in \cite{Lubow2010} that is corrected in \cite{Lubow2012}.}, where $\chi = 2.08 q^2 \Omega_{\rm b} r_{\rm res}$ \citep{Lubow1991, Ogilvie07}.

We use the boundary conditions that $dE/dr=0$ at the disk inner and outer radius and the initial condition is given by
\begin{equation}
   E(r,0) = \cos{\left[ \frac{\pi}{2} \left( \frac{r_{\rm out} - r}{r_{\rm out} - r_{\rm in}}  \right) \right]}.
\end{equation}
We solve the time dependent equation~(\ref{eq:EQ1}) and take the steady state solution reached at a time of $t=200\,P_{\rm b}$. In the steady state, the disk approaches an exponentially growing eigenmode in which the eccentricity versus radius, normalized by the eccentricity at the outer radius, is constant and disk undergoes apsidal precession at a constant rate.

We assume that the time dependence of the eccentricity has the form $E\propto \exp (i \omega t)$. The eccentricity  growth  rate and the precession rate are then given by
\begin{equation}
i \omega=    \frac{\dot E}{E}=\frac{\dot e}{e}+i\dot \varpi.
\end{equation}
The eccentricity growth rate is $-\Im (\omega)$ and the precession rate is $\Re (\omega)$. These quantities are fixed with radius and time once the disk reaches an eigenmode.

We first consider the effect of changing the outer disk radius. 
Fig.~\ref{fig:analytic} shows eccentricity growth rate and the precession rate with varying outer disk radius for four different binary mass ratios for two disk aspect ratios. The line with mass ratio $q=0.1$ and $H/r=0.02$ was presented in \cite{Vallet2026}. For these parameters, the apsidal precession is most negative when the disk outer radius is close to the resonance location. The eccentricity growth rate peaks just outside of the resonance location. For a larger $H/R=0.03$, the behavior is qualitatively similar, except the lowest apsidal precession rate is close to zero.

The behavior for $q=0.2$ is similar to $q=0.1$. However, the resonance location is farther out relative to the Roche lobe radius. The minimum apsidal precession rate is negative for both values of $H/r$.  Even for large disk truncation radius, the apsidal precession does not quite become positive for large truncation radius. It can only be positive for smaller outer radius. 

On the other hand, for the high mass ratios ($q=0.4$ and $q=0.6$) the resonance location is outside of the disk. The outer radius where the precession becomes positive would be far beyond the Roche lobe radius. Therefore,  positive precession requires the disk to be truncated at smaller radius for high binary mass ratio. While the SPH simulations in Section~\ref{sec:hydro} showed that the timescale for eccentricity growth was too long for growth for the parameters we used with $q=0.6$, these results suggest that eccentricity growth and prograde apsidal precession may be possible for $q=0.6$ (and larger) for a limited range of outer disk radii. Eccentricity growth has been seen in 2D hydrodynamical grid code simulations up to $q=1$ \citep{Kley2008}.

\subsection{Effect of a density pile up}

\begin{figure}
    \centering
    \includegraphics[width=\linewidth]{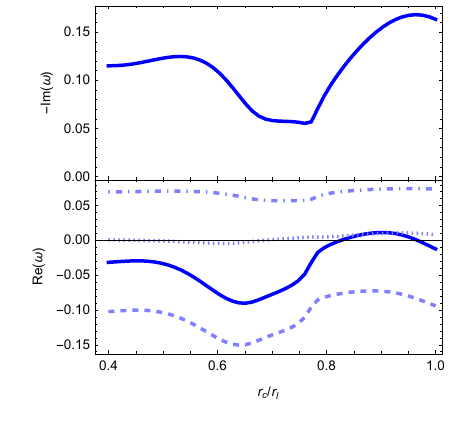}
    \caption{The eccentricity growth rate (upper) and the precession rate (lower) as a function of the disk outer radius for the
linear models with surface density profile  given in equation~(\ref{density}) with $C=5$, $\sigma=0.06$ and a varying center for the Gaussian peak, $r_{\rm c}$. The mass ratio is $q=0.4$, $H/r=0.02$, and $r_{\rm out}=0.85\,r_{\rm l}$. In the lower panel, the dot-dashed line shows the contribution from the dynamical gravitational term (equation~(\ref{dyn})), the dashed line shows the contribution from the $de/dr$ term (equation~(\ref{egrad})) and the dotted line shows the contribution from the pressure terms (equation~(\ref{pressure})).}
    \label{fig:gaussian}
\end{figure}

To isolate how the mass distribution in the disk affects the precession, we introduce a density perturbation. We take
\begin{equation}
    \Sigma \propto \left[\frac{1}{r^{1/2}}+C \exp \left(-\frac{(r-r_{\rm c})^2}{\sigma^2} \right) \right],
    \label{density}
\end{equation}
where $C=5$, $\sigma=0.05$ and we vary the radius on which the Gaussian is centered, $r_{\rm c}$. The density is distributed between $r_{\rm in}=0.01\,a_{\rm b}$ and $r_{\rm out}=0.85\, r_{\rm l}$.

Fig.~\ref{fig:gaussian} shows the effect of moving the Gaussian density bump. When the bump is at small radius (where it does not affect the density distribution in the resonance), the location does not change the eigenmode. For this mass ratio of $q=0.4$, the precession is retrograde. As the density bump moves outwards, the growth rate drops off and the precession becomes more retrograde. As the bump moves beyond the resonance location, the eccentricity growth again increases and the precession becomes prograde. This shows that the precession direction is very sensitive to the distribution of material within the disk.  A build up of material in the outer parts of the disk can lead to more prograde apsidal precession, even for high mass ratio systems.  Relatively small changes in the radial mass distribution can reverse the direction of precession.

The apsidal precession rate can be written as
\begin{equation}
    \Re (\omega) = \frac{I_{\rm egrad}+I_{\rm pressure}+I_{\rm dyn}}{I}
\end{equation}
\citep[e.g.][]{Teyssandier2017},
where
\begin{equation}
    I=2\int_{r_{\rm in}}^{r_{\rm out}} r^3 \Sigma \Omega |E|^2\,dr.
\end{equation}
The contribution from the dynamical gravitational term is
\begin{equation}
    I_{\rm dyn}= \int_{r_{\rm in}}^{r_{\rm out}}\frac{\Sigma q r^5\Omega^2}{2a_{\rm b}^2}b_{3/2}^{(1)} |E|^2\, dr,
    \label{dyn}
\end{equation}
where $b^{(1)}_{3/2}(r/a_{\rm b})$ is the Laplace coefficient.
The term that includes the gradient of the eccentricity is
\begin{equation}
    I_{\rm egrad}=-\int_{r_{\rm in}}^{r_{\rm out}}\left(2-\frac{1}{\gamma}\right)Pr^3\left|\frac{\partial E}{\partial r}\right|^2 \, dr.
    \label{egrad}
\end{equation}
The contribution from pressure is
\begin{equation}
    I_{\rm pressure}=\int_{r_{\rm in}}^{r_{\rm out}} \left[ \left(4-\frac{3}{\gamma}\right) r \frac{dP}{dr}+3\left(1+\frac{1}{\gamma}\right)P \right] r|E|^2\, dr
\label{pressure}
\end{equation}
The lower panel of Fig.~\ref{fig:gaussian} shows each of the contributions to the precession rate. The dynamical gravitational term always drives prograde precession. The pressure term is a relatively small contribution. The dominant negative part is a result of the gradient of the eccentricity.  This helps to explain why the SPH simulations show prograde precession, because the eccentricity gradient is smoothed out and therefore the dominant negative contribution is weakened.

\section{Conclusions}
\label{concs}

Positive superhumps in CVs are generally attributed to prograde apsidal precession of an eccentric accretion disk driven by the 3:1 resonance. Because the resonance lies outside the tidal truncation radius for binaries with $q >q_{\rm crit} \approx 0.33$, it has long been assumed that eccentric disks and positive superhumps can only occur in low mass ratio systems. However, several CVs with high mass ratios have been observed to show positive superhumps (see Table~\ref{table}).

We have shown that eccentricity growth can occur even when the 3:1 resonance radius lies outside the disk. Our three–dimensional hydrodynamic simulations demonstrate strong eccentricity growth for $q=0.4$, indicating that the finite width of the resonance can allow it to influence the outer parts of the disk. In all of our simulations the disk undergoes prograde apsidal precession, which would produce positive superhumps.

Using linear analysis we find that the direction of apsidal precession is highly sensitive to the disk surface density structure. In particular, the outer disk radius and the radial distribution of mass can determine whether the disk precesses in a prograde or retrograde direction. The relatively narrow disks produced in SPH simulations favor prograde precession, while broader disks with stronger eccentricity gradients can undergo retrograde precession. These results suggest that the positive superhumps observed in some high mass ratio CVs can be explained by eccentric disks whose structure favors prograde apsidal precession, even though the resonance lies formally outside the disk.

%% Please use the acknowledgment and contribution environments. This will 
%% be anonomyized when the "anonymous" style option is used. 
\begin{acknowledgments}
We thank an anonymous referee for useful comments that improved the paper. 
%We thank all the people that have made this AASTeX what it is today.  This
%includes but not limited to Bob Hanisch, Chris Biemesderfer, Lee Brotzman,
%Pierre Landau, Arthur Ogawa, Maxim Markevitch, Alexey Vikhlinin and Amy
%Hendrickson. Also special thanks to David Hogg and Daniel Foreman-Mackey
%for the new {\tt\string modern} style design. Considerable help was provided via bug
%reports and hacks from numerous people including Patricio Cubillos, Alex
%Drlica-Wagner, Sean Lake, Michele Bannister, Peter Williams, Jonathan
%Gagne, Arthur Adams, Nicholas Wogan, Aaron Pearlman, Jeff Mangum, and Mark Durre.
\end{acknowledgments}

\software{
Phantom \citep{Price2018} 
%Splash \citep{Price2007}
          }


\bibliography{NegSH25}{}
\bibliographystyle{aasjournalv7}



\end{document}